\documentclass{aa}
\usepackage{times}
\usepackage{graphics}

\begin{document}

\thesaurus{05(10.07.3 M22; 10.07.3 M55; 08.01.1; 08.02.7)}

\title{Str\"omgren photometry in globular clusters : M55 \& M22 
\thanks{Based on data collected at the European Southern Observatory,
	 La Silla, Chile}}

\author{P. Richter \inst{1}
        \and
        M. Hilker \inst{2} 
        \and
        T. Richtler \inst{1}} 

\offprints{prichter@astro.uni-bonn.de}

\institute{
Sternwarte der Universit\"at Bonn, Auf dem H\"ugel 71, 53121 Bonn, Germany
\and
Departamento de Astronom\'\i a y Astrof\'\i sica, P.~Universidad Cat\'olica,
Casilla 104, Santiago 22, Chile
}

\date{Received 6 April 1999/ Accepted 13 July 1999}

\titlerunning{Str\"omgren photometry in globular clusters : M55 \& M22}
\authorrunning{P. Richter et al.}
\maketitle

\begin{abstract}
We present Str\"omgren CCD photometry for the two
galactic globular clusters M55 (NGC\,6809) and M22 (NGC\,6656). 

We find average Str\"omgren metallicities of $-1.71\pm 0.04$ dex for M55
and $-1.62\pm 0.08$ dex for M22.
The determination of metal
abundances in cluster giants with the Str\"omgren $m_1$ index in comparison 
with spectroscopic data from Briley et al. (1993) and Norris \&
Freeman (1982, 1983) shows that M55 and M22 have different
distributions of cyanogen strengths.
In M55, no CN abundance variations are visible among the giant-branch stars.
In striking contrast, a large dispersion of cyanogen strengths is seen in M22. 
For M22 we find patchily distributed variations in the foreground reddening
of $\Delta E(B-V) \approx 0.07$,  
which explain the colour dispersion among the
giant-branch stars. There is no evidence
for a spread in iron within M22 since the variations in $m_1$ are 
dominated by the large range in CN abundances, as already found
by Anthony-Twarog et al. (1995).
The difference between M55 and M22 may resemble the difference in integral
CN band strength between 
M31 globular clusters and the galactic system. 

The colour-magnitude diagram of M55 shows the presence of a
population of 56 blue-straggler stars that are more centrally
concentrated than the red giant-branch stars.

\keywords{globular clusters: individual: M22 and M55  
           -- stars: abundances -- blue stragglers}

\end{abstract}


\section{Introduction}
The phenomenon of chemical inhomogeneity within galactic globular clusters
is still not clearly understood (see reviews by Suntzeff 1993 and Kraft 1994). 
While there 
is only $\omega$ Centauri and 
perhaps also M22 which show variations in their iron abundances,
many globular clusters have 
variations in elements like C, N and O (e.g. Hesser 1976).
However, other clusters seem to be chemically very homogeneous.

In searching for explanations of abundance variations within globular clusters
there are basically two possibilities: primordial variations and inhomogeneities
caused by stellar evolution during the giant branch (GB) or later phases.
Studies of the abundances of CNO elements in globular clusters have not led to
clear results yet. On the one hand, most of the cyanogen (CN) variations can be
successfully explained by processes during the CNO cycle (Kraft 1994).
On the other hand, CN variations in globular clusters also have been
found in main-sequence stars (Suntzeff 1989, Briley et al. 1991), which
points to primordial inhomogeneities. Therefore it is very interesting to
make a large census of CN band strenghts in both giants and main-sequence stars.

Since heavy elements like iron are not synthesized in present globular-cluster
stars, variations of the iron abundance among cluster stars are 
believed to be of primordial origin. However, in the case of the
 extraordinarily
massive globular cluster $\omega$ Centauri two other explanations 
are under discussion. First, there might have been
secondary star formation within the cluster from enriched gas that was not
blown out of the cluster during its formation (e.g. Norris et al. 1996). 
Second, $\omega$ Centauri might be the result of the merging of two clusters
with different metallicities (Searle 1977, Icke \& Alcaino 1988, Norris et al.
1997).

The inhomogeneity in $\omega$ Centauri is visible in the colour-magnitude diagram
(CMD) by way of a significant colour dispersion among the giant-branch stars 
(e.g. Persson et al. 1980). Since a broad colour dispersion is visible on the giant 
branch of the smaller cluster M22 as well 
(as mentioned already by Arp \& Melbourne 1959), this cluster
has also been regarded as a candidate for a primordial abundance spread.
However, all the iron-rich stars of the sample (DDO survey) by Hesser et al.
(1977) have been identified as non-members in later studies (Lloyd Evans 1978,
Peterson \& Cudworth 1994). Moreover, in more recent spectroscopic
investigations (Lehnert et al. 1991, Brown \& Wallerstein 1992) measurable
variations in iron are not confirmed, whereas variations in CN are clearly
detectable (Norris \& Freeman 1982, 1983; Brown et al. 1990).
Lehnert et al. (1991) give an upper limit for a
possible metallicity spread of $\Delta$[Fe/H]$=0.2$ dex.
The origin of the colour dispersion in M22 might rather be explained by
differential foreground reddening. A spectroscopic analysis by Crocker 
(1988) and polarisation measurements by Minniti et al. (1990, 1992) constrain
the reddening variations to be less than 0.08 mag. Bates et al. (1992),
investigating the region of M22 with IRAS data, give a value of $\Delta E(B-V)
\approx$ 0.05 for the cluster field.

Several studies have shown that the Str\"omgren $m_1$ index ($m_1$ is the
difference between the colour indices $(v-b)$ and $(b-y)$) ist not only
a good indicator for the mean metallicity of late type stars, but is also 
sensitive for CN abundances (e.g. Bell \& Gustafsson 1978).
CCD Str\"omgren photometry offers the possibility to measure metallicities
and cyanogen strengths of many giants in globular clusters simultaneously.
It is therefore a appropriate tool for adressing the question of metallicity/CN
band distribution with a much larger sample of stars than spectroscopy. 

Anthony-Twarog et al. (1995) already applied the Str\"omgren system to M22. They
found that the m$_1$ index is indeed closely correlated with CN-band strengths and
that CN-variations are found all over the giant branch.  

The intention of the present paper is to make a differential comparison between
M55 and M22 and to investigate the influence of the differential reddening in M22.
We therefore re-investigated the anomal abundance anomalies of M22 with $vby$ photometry
and combine our results with spectroscopic measurements of Norris \& Freeman
(1982, 1983; hereafter refered to as NF). Moreover, we discuss photometric results for M55, for which
spectroscopic measured CN abundances are also available (Smith \& Norris 1982;
Briley et al. 1993).
M55 is known to be rather monometallic (Zinn \& West 1984)
and it is therefore an excellent comparison object in order to investigate
CN abundances of two galactic globular clusters with different 
chemical properties. 

Sect.~2 will give an overview of the observations and the data reduction,
Sect.~3 and Sect.~4 will present the results and their discussion for the two
clusters. In Sect.~5 we summarize our results and give options for future
works.

\begin{table}[t]
\caption[]{\label{fields}Centre positions of CCD pointings for M22 and M55 
fields}
\begin{flushleft}
\begin{tabular}{lccc}
\hline\noalign{\smallskip}
Object & Field & $\alpha_{2000}$ & $\delta_{2000}$\\
\noalign{\smallskip}
\hline\noalign{\smallskip}
M22 & A & 18:36:35.2 & $-$23:56:36 \\
    & B & 18:36:35.2 & $-$23:52:06 \\
    & C & 18:36:15.1 & $-$23:56:36 \\
    & D & 18:36:15.1 & $-$23:52:06 \\
M55 & E & 19:40:18.9 & $-$30:57:08 \\
    & F & 19:40:09.9 & $-$30:54:52 \\
    & G & 19:39:49.9 & $-$30:54:51 \\
    & H & 19:39:49.9 & $-$30:59:26 \\
    & I & 19:40:09.7 & $-$30:59:26 \\
\noalign{\smallskip}
\hline
\end{tabular}
\end{flushleft}
\end{table}

\begin{table*}[t]
\caption[]{\label{log} Log of CCD observations}
\begin{flushleft}
\begin{tabular}{lllllllllllll}
\hline\noalign{\smallskip}
Object & Field & Filter & Expos. time [s] & Date        &  & Object & Field & Filter & Expos. time [s] & Date \\
\noalign{\smallskip}
\hline\noalign{\smallskip}
M22    & A     & $y$    & 30,150,150      & 1995 Apr 23 &  & M55    & E     & $y$    & 30,120          & 1995 Apr 21 \\
       &       & $b$    & 60,270,270      &             &  &        &       & $b$    & 60,180          &             \\
       &       & $v$    & 120,600,600     &             &  &        &       & $v$    & 120,600         &             \\
       & B     & $y$    & 30,90,120       &             &  &        & F     & $y$    & 70              & 1995 Apr 22 \\
       &       & $b$    & 60,180,240      &             &  &        &       & $b$    & 120             &             \\
       &       & $v$    & 120,420,540     &             &  &        &       & $v$    & 240             &             \\
       & C     & $y$    & 30,120,120      & 1995 Apr 24 &  &        & G     & $y$    & 70              &             \\
       &       & $b$    & 60,240,240      &             &  &        &       & $b$    & 120             &             \\
       &       & $v$    & 120,480,480     &             &  &        &       & $v$    & 240             &             \\
       & D     & $y$    & 40              &             &  &        & H     & $y$    & 70              &             \\
       &       & $b$    & 60              &             &  &        &       & $b$    & 120             &             \\
       &       & $v$    & 120             &             &  &        &       & $v$    & 240             &             \\
       &       &        &                 &             &  &        & I     & $y$    & 70              &             \\
       &       &        &                 &             &  &        &       & $b$    & 120             &             \\
       &       &        &                 &             &  &        &       & $v$    & 240             &             \\
\noalign{\smallskip}
\hline
\end{tabular}
\end{flushleft}
\end{table*}


\section{Observations and data reduction}

\begin{table}[b]
\caption[]{Coefficients for the calibration equations}
\begin{flushleft}
\begin{tabular}{ccccc}
\hline\noalign{\smallskip}
Colour & A & B & C & RMS\\
\noalign{\smallskip}
\hline\noalign{\smallskip}
$y$ & 4.080 & 0.136 & $-$0.022 & 0.015 \\
$b$ & 3.570 & 0.180 & $-$0.053 & 0.015 \\
$v$ & 4.010 & 0.275 & 0.014  & 0.017 \\
\hline
\end{tabular}
\end{flushleft}
\end{table}

The observations have been performed in the nights  21.-24. April 1995
with the Danish 1.54m telescope at ESO/La Silla. The CCD in use
was a Tektronix chip with 1024$\times$1024 pixels.
The $f$/8.5 beam of the telescope provides a scale of $15\farcs7$/mm,
and with a pixel size of 24 $\mu$m the total field is $6\farcm3 \times 
6\farcm3$. A total number of 66 images (22 in each colour) has been taken
during the 4 nights through Str\"omgren $vby$ filters.
Table 2 shows the position of the CCD fields for the
two clusters while Table 3 contains information about the observations. 
Furthermore, 16 frames containing 17 standard stars from J{\o}nch-S{\o}rensen 
(1993,1994) were obtained as well, using 14 E region stars from the publication
of 1993 and 3 faint stars from the one of 1994. All frames were
obtained under good seeing conditions (FWHM of $1\farcs0 - 1\farcs3$).

The CCD frames were processed with the standard IRAF routines, instrumental
magnitudes were derived using DAOPHOT and ALLSTAR (Stetson 1987, 1992).
Calibration equations have been determined for $V(=y)$, $b$ and $v$,
taking the according airmasses $X$ into consideration:

\begin{flushleft}
$y_{ins} = y_{st} + A_{y} + B_{y}\cdot X_{y} + C_{y}\cdot (b-y)_{st}$\\
\vspace{0.2cm}
$b_{ins} = b_{st} + A_{b} + B_{b}\cdot X_{b} + C_{b}\cdot (b-y)_{st}$\\
\vspace{0.2cm}
$v_{ins} = v_{st} + A_{v} + B_{v}\cdot X_{v} + C_{v}\cdot (v-b)_{st}$\\
\end{flushleft}

The coefficients for the calibration equations are given in Table 3.
The calibration for $(b-y)$ and $m_1$ have been calculated from the values of 
$y,b,v$, respectively. After the photometric reduction and after matching all frames,
upper limits for the total photometric errors are $0.015$ mag for $V$, $0.019$ mag for
$(b-y)$ and $0.029$ mag for $m_1$.

\begin{figure}
\resizebox{\hsize}{!}{\includegraphics{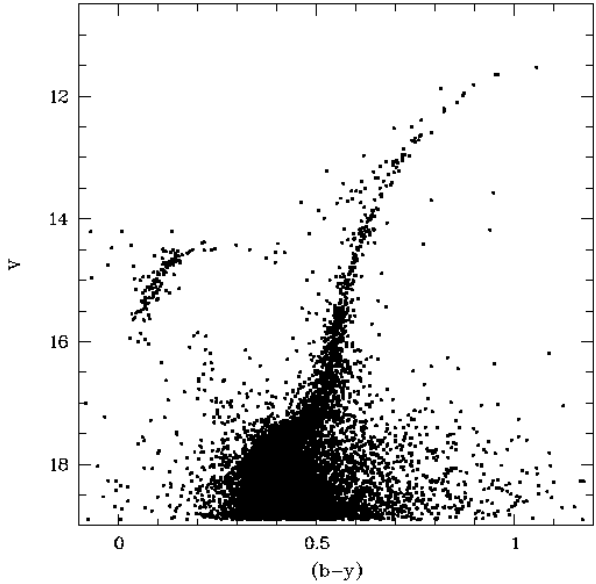}}
\caption{The colour-magnitude diagram of M55. It is cleaned
by error cuts ($\sigma_V < 0.2$ mag) }
\end{figure}

\section{M55}
\subsection{The colour-magnitude diagram}
Fig.\,1 shows the CMD of M55. From the original data set of 17269 stars,
all objects within an inner radius of 320$\farcs0$ have been taken.
We selected according to a maximum error in $V$ of $\sigma_V < 0.2$ mag 
for the stars brighter than $V = 20.5$.
The narrow giant branch and the well defined blue horizontal branch (BHB)
show the high internal precision of the
selected subsample for the bright stars in M55.
Below $V = 17.0$ the photometric errors increase quickly and produce 
the large scatter near the turn-off point.

A number of blue-straggler stars populate the
CMD in the region between the horizontal branch and the turn-off point. 
These objects will be discussed in the appendix.

\begin{figure}
\resizebox{\hsize}{!}{\includegraphics{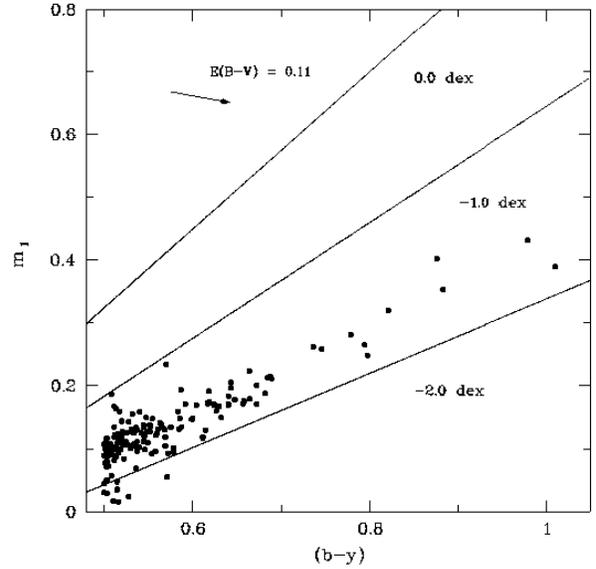}}
\caption{The $m_1,(b-y)$ diagram of M55 shows a sharp
sequence of giant branch stars, in striking contrast to M22, where the
distribution is much broader. The diagram has been reddening corrected
with $E(B-V)=0.11$. The indicated lines of constant metallicity
are taken from the calibration of Hilker (priv. comm.)}
\end{figure}

\subsection{The $m_1,(b-y)$ diagram}
In the $m_1,(b-y)$ diagram of M55 (Fig.~2) stars with $V<16$ and $(b-y)>0.5$
have been plotted. The $m_1,(b-y)$ diagram shows a well defined sequence 
of the giant-branch stars and there is no evidence for a colour spread
larger than the photometric errors.
The diagram is reddening corrected with $E(B-V)=0.11$. We obtain this value 
from our photometric data, as we show in the following section.

Through the correlation between [Fe/H], $m_1$ and $(b-y)$ 
the $m_1,(b-y)$ diagram can be used to derive metallicities
for individual giant-branch stars in globular clusters (Grebel \& Richtler 1992).
A new calibration of the $m_1,(b-y)$ diagram 
for the metallicity of red giants
is given by Hilker (priv. comm.), in Fig.~2 indicated for 
the metallicities of $0.0$,$-1.0$ and $-2.0$ dex.
It is based on a sample of 60 stars with known
Str\"omgren colours and spectroscopically determined [Fe/H]-values
in the range $-2.4$ dex $\le$ [Fe/H] $\le$ $-0.6$ dex. 
36 stars are taken from the work of
Anthony-Twarog \& Twarog (1998), 17 from a sample
in $\omega$ Centauri and 7 stars of
M22 from this work. The metallicity then is
given by:
\begin{displaymath}
{\rm [Fe/H]} =\frac{m_{1,0}- 1.253 \pm 0.062 \cdot (b-y)_0 + 0.303 \pm 0.046}
{ 0.331 \pm 0.045 \cdot (b-y)_0 - 0.026 \pm 0.033}
\end{displaymath}
For stars more metal rich than [Fe/H] = $-0.6$ dex
the new calibration has been adjusted to the older one of Grebel \&
Richtler (1992).
Values for [Fe/H] derived by this method are sensitive to the 
adopted reddening. 
However, the CMD of a globular cluster also contains information
about its reddening and metallicity.
Gratton \& Ortolani (1989) find the following correlation 
between [Fe/H] and the colour of the giant branch in globular clusters 
at the level of 
the horizontal branch, $(B-V)_{\rm 0,g}$ :
\begin{displaymath}
{\rm [Fe/H]} = 2.85 \pm 0.37 \cdot (B-V)_{\rm 0,g} - 3.76 \pm 0.31
\end{displaymath}
For these two equations reddening and metallicity are
correlated in different ways, so that with the combination of both an unambiguous
result for [Fe/H] {\it and} $E(B-V)$ can be obtained.

For M55 we find $(b-y)_{\rm 0,g} = 0.60 \pm 0.03$, equivalent to
$(B-V)_{\rm 0,g} = 0.83 \pm 0.06$ (Grebel \& Roberts, priv. comm.).
Following Hilker's calibration of 
the $m_1,(b-y)$ diagram for stars with $(b-y) \ge 0.6$ we 
find a reddening of $E(B-V)=0.11 \pm 0.02$ and  
a mean Str\"omgren metallicity of 
[Fe/H]$=-1.71 \pm 0.04$ dex for M55.
This result is, within the error range, in good
agreement with the value of 
Zinn \& West (1984; $-1.82 \pm 0.15$ dex).

\begin{table}[t]
\caption[]{CN strengths and Str\"omgren colours of red giants in M55. The values of
$S$(3839) have been taken from Briley et al. (1993)}
\begin{flushleft}
\begin{tabular}{lcccrr}
\hline\noalign{\smallskip}
Nr.$_{\rm Lee}$ & $V$ & $(b-y)$ & $m_{1}$ & $\Delta m_{1}$ & $S(3839)$ \\
\noalign{\smallskip}
\hline\noalign{\smallskip}
 1317  & 11.949  & 0.874 &  0.232 & $-$0.026 & 0.01 \\
 1319  & 13.089  & 0.707 &  0.151 & $-$0.005 & 0.16 \\
 1401  & 12.103  & 0.856 &  0.265 & $+$0.018 & 0.20 \\
 1404  & 13.857  & 0.644 &  0.116 & $-$0.002 & 0.08 \\
 1418  & 13.042  & 0.717 &  0.153 & $-$0.010 & 0.12 \\
 1518  & 11.639  & 0.953 &  0.385 & $+$0.079 & 0.10 \\
 1523  & 13.158  & 0.702 &  0.153 & 0.000    & 0.11 \\
 2307  & 12.628  & 0.766 &  0.195 & $+$0.002 & 0.31 \\
 2318  & 12.948  & 0.725 &  0.162 & $-$0.005 & 0.19 \\
 2437  & 11.813  & 0.898 &  0.303 & $+$0.030 & 0.09 \\
 2502  & 11.653  & 0.959 &  0.336 & $+$0.026 & 0.18 \\
 2510  & 13.076  & 0.701 &  0.153 & 0.000    & 0.21 \\
 2517  & 13.040  & 0.704 &  0.144 & $-$0.011 & 0.09 \\
 3321  & 11.988  & 0.871 &  0.248 & $-$0.007 & 0.09 \\
 3402  & 13.548  & 0.601 &  0.089 & $-$0.007 & 0.07 \\
 3409  & 12.854  & 0.721 &  0.180 & $+$0.015 & 0.10 \\
 3509  & 12.247  & 0.822 &  0.241 & $+$0.014 & 0.02 \\
 3516  & 13.498  & 0.661 &  0.143 & $+$0.014 & 0.20 \\
 3519  & 13.849  & 0.656 &  0.084 & $-$0.041 & 0.24 \\
 3526  & 12.860  & 0.721 &  0.188 & $+$0.023 & 0.12 \\
 4323  & 12.974  & 0.734 &  0.154 & $-$0.018 & 0.20 \\
 4403  & 12.743  & 0.749 &  0.154 & $-$0.028 & 0.13 \\
 4422  & 12.394  & 0.765 &  0.198 & $+$0.006 & 0.08 \\
 4431  & 13.386  & 0.669 &  0.154 & $+$0.020 & 0.22 \\
 4501  & 13.403  & 0.692 &  0.113 & $-$0.034 & 0.20 \\
 4505  & 11.532  & 1.057 &  0.414 & $+$0.045 & 0.28 \\
 4523  & 12.499  & 0.741 &  0.162 & $-$0.015 & 0.26 \\
\noalign{\smallskip}
\hline
\end{tabular}
\end{flushleft}
\end{table}

\begin{figure}
\resizebox{\hsize}{!}{\includegraphics{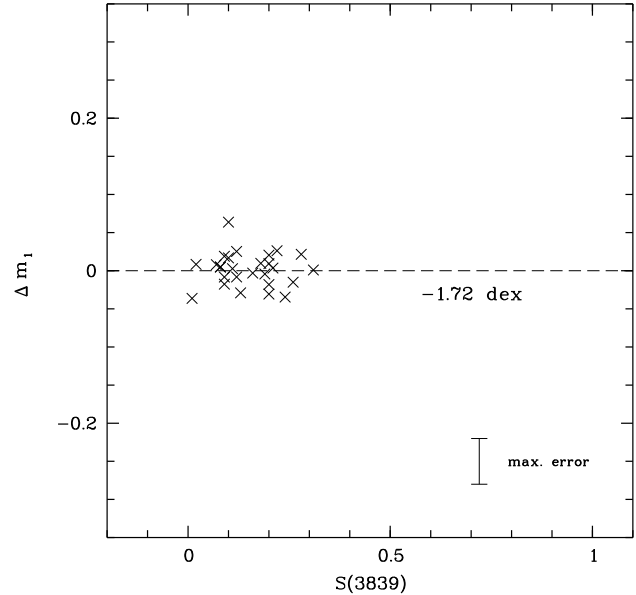}}
\caption{The dispersion of the $m_1$ values does not
correspond to the CN abundances in M55. The errorbar in the lower right corner
indicates the photometric error in $m_1$}
\end{figure}

\subsection{CN abundances}

Richtler (1988) found (on the basis of aperture photometry) that the giant
branch in the M55 $m_1,(b-y)$ diagram splits up in two sub-branches where the 
corresponding metallicities anticorrelate with the CN band strengths given
by Smith \& Norris (1982). Since in other clusters an anticorrelation between
C and N is well established (e.g. Suntzeff 1993), the suspicion was that 
a continous absorption of CO, which affects the Str\"omgren $v$-filter, 
could be responsible for this effect. 

For comparison, we took 27 stars from the sample of Briley et al. (1993)
with known CN abundances (Table 4), identified by the identification numbers
from Lee (1976). This sample also includes results from the
measurements of Smith \& Norris (1982).
The range of cyanogen strengths  
is given by $0.01 < S(3839) < 0.31$ (Briley et al. 1993),
where $S(3839)$ is the spectroscopic index for the CN band
beginning at 3883\,\AA\ (see Norris \& Freeman 1982 for a detailed description).   
In our $m_1,(b-y)$ diagram of M55 (Fig.~2) the scatter in $m_1$ is small
and therefore no signs for significant CN variations are visible.
The mean Str\"omgren metallicity of the selected 27 stars is $-1.72$ dex, 
derived with the calibration of Hilker.
We define $\Delta m_1$ to be the vertical distance from 
the position of an individual star in the $m_1,(b-y)$ diagram to the calibration
line of $-1.72$ dex. In Fig.\,3, $\Delta m_1$ has been plotted versus
the cyanogen index $S(3839)$. No correlation between the $\Delta m_1$
values and the cyanogen strengths can be seen. The variations of the
$\Delta m_1$ values are of similar size as the photometric errors of
the $m_1$ index, so that the $m_1$-CN anticorrelation found by Richtler (1988)
might be below our detection limit. 
However, our measurements give no indication for a significant abundance
spread in CN in the cluster giants of M55.

The comparison between the photometric study of Richtler (1988) and
the present data reveals only slight differences in the measured Str\"omgren 
colours. From a comparison of 7 stars we find that our values
of $V$, $(b-y)$ and $m_1$ show mean deviations of $\Delta V =+0.018$, $\Delta (b-y)=-0.002$ 
and $\Delta m_1 =+0.054$ (in the sense of $\Delta$colour $=$ colour$_{\rm present\, data} - 
$colour$_{\rm Richtler}$).
It is most likely that these shifts reflect 
the usage of different standard stars (see Richtler 1988).
Anyway, the differences are of no importance for the qualitative comparison
described above.

\begin{figure}
\resizebox{\hsize}{!}{\includegraphics{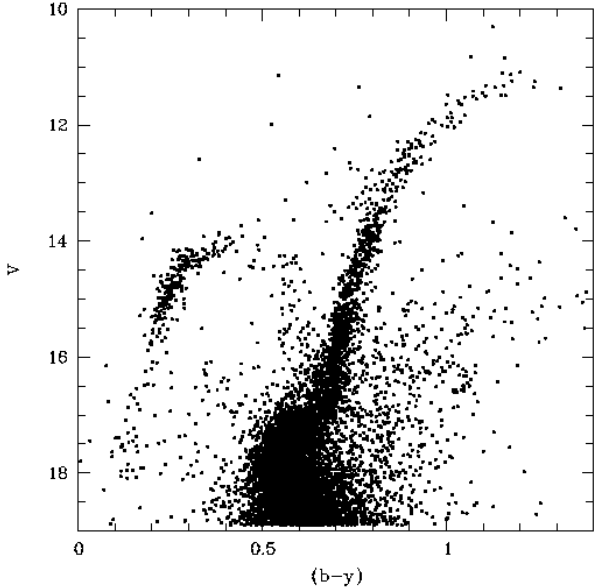}}
\caption{The colour-magnitude diagram of M22 reveals
a significant colour dispersion among the red giant branch}
\end{figure}

\section{M22}
\subsection{The colour-magnitude diagram}
The CMD of M22 is presented in Fig.~4. It contains all stars  
brighter than $V=19.0$ and with $57\farcs0 <$ $r$ $< 302\farcs0$. 
The stars from the inner part have been omitted because of the high star
density in the center of M22 and the resulting large photometric errors. 
The shown sample has been cleaned by error selection as done for M55.

\begin{figure}
\resizebox{\hsize}{!}{\includegraphics{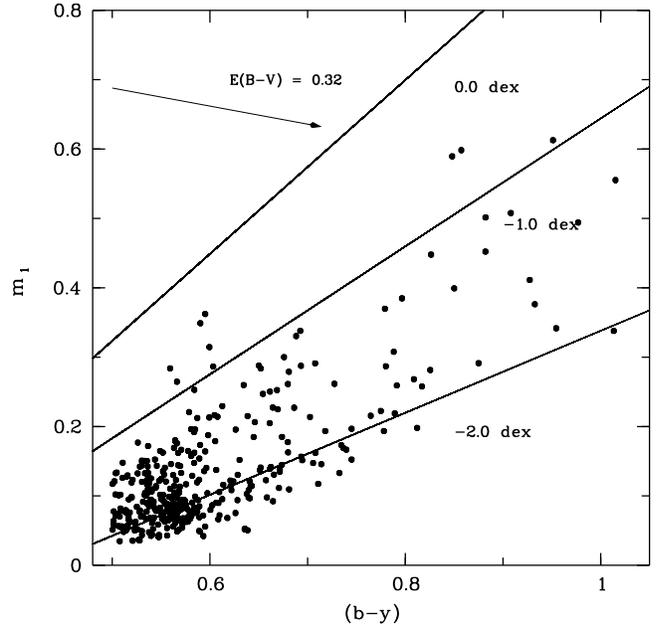}}
\caption{The $m_1,(b-y)$ diagram of M22 shows a large scatter which
can be explained by abundance variations as well as by reddening effects.
The diagram has been reddening corrected with $E(B-V)=0.32$, as given
by Alcaino \& Liller (1983)}
\end{figure}

The basic features of the red giant branch (RGB) and the blue horizontal branch
(BHB) are well defined. In striking difference to M55 is the large colour
spread among the red giant branch, as already mentioned by Arp \& Melbourne
(1959). A similar spread is also visible on the BHB, which suggests
that differential  reddening is the dominating effect for the origin
of the colour dispersion in the CMD. However, since a contribution
to the colour range by a possible metallicity spread in M22 can not
be excluded, the CMD alone can not clarify the situation.

In the following, we select red giants by defining the ridges of the red giant
branch by two parallel curves which at $V = 14$ are separated in colour by 
$0.08$ mag.

\subsection{The $m_1,(b-y)$ diagram}
Fig.\,5 shows the $m_1,(b-y)$ diagram of M22. Selected are stars from
the red giant branch with $V<16.0$ and $(b-y)>0.5$. 
In sharp contrast to M55, the distribution of stars
does not exhibit a well defined straight line but shows a strongly
scattered distribution, so that an accurate determination of a mean
cluster metallicity with the method described in Sect.\,3.2 is
not possible without knowing the reasons for the colour spread.
 
Previous measurements indicate that the foreground reddening in
the direction to M22 is $\ge 0.30$ mag (e.g. Alcaino \& Liller 1983, Crocker 1988).
In addition, recent spectroscopic studies in M22 
lead to a mean cluster metallicity
of $-1.48$ dex (Carretta \& Gratton 1997).
The $m_1,(b-y)$ diagram shown in Fig.\,5 is reddening corrected with $E(B-V)=0.32$,
equivalent to the value given by Alcaino \& Liller (1983). 
Using Hilker's calibration, some of the giant branch stars with $(b-y) \ge 0.6$ lie
in a range higher than $-1.2$ dex, so that, with respect to the
expected mean metallicity, the giants seem to
be scattered to higher metallicities. 
Since any additional reddening would further increase the metallicity,
differential reddening is not a likely candidate in explaining the broad
distribution. Moreover, we demonstrate directly  that differential reddening
does not play a role but that the CN band strengths are responsible.

\begin{figure}
\resizebox{0.95\hsize}{!}{\includegraphics{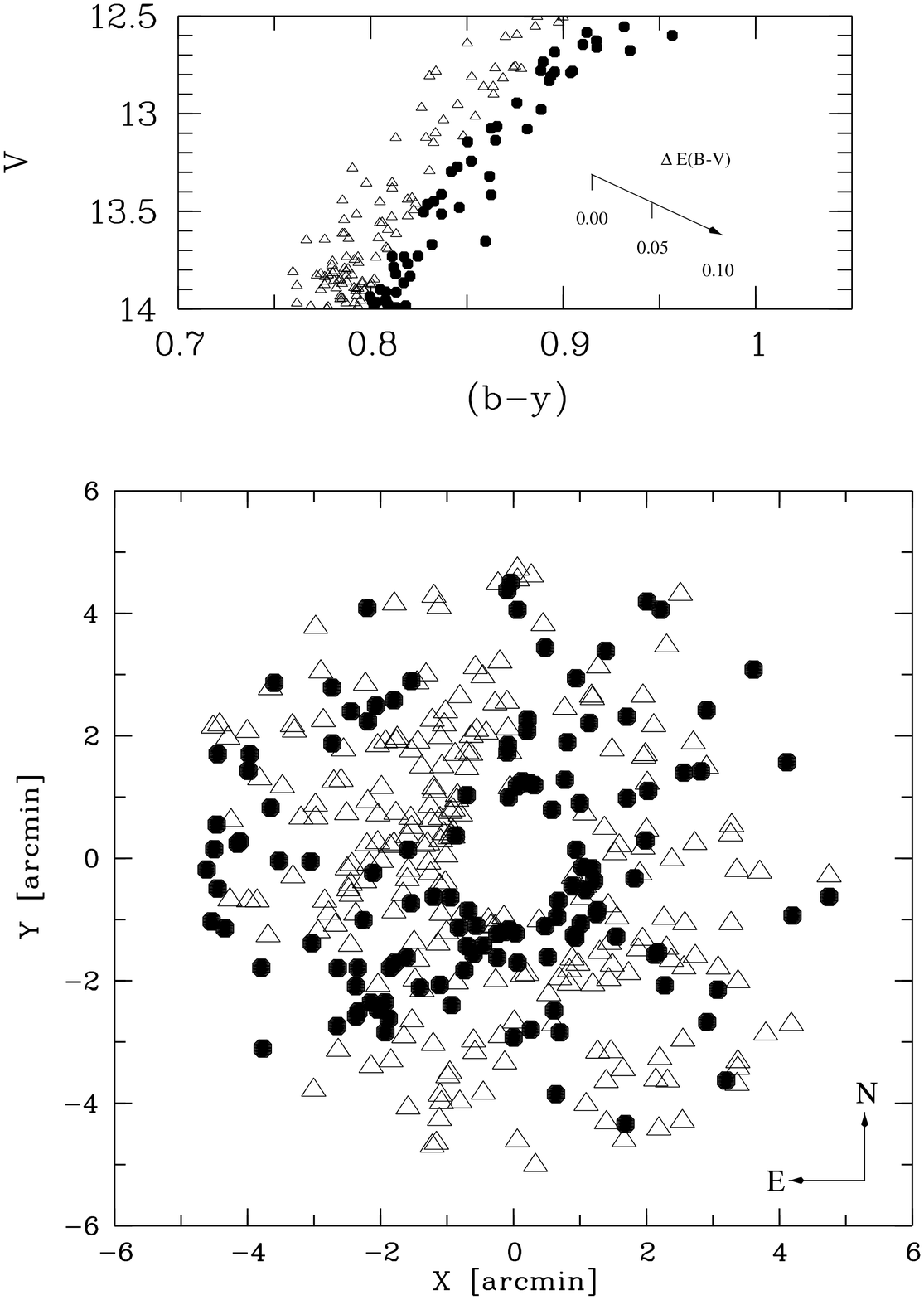}}
\caption{Upper panel: The wide RGB of M22, splitted in the redder
giants (filled dots) and the bluer ones (open triangles). Lower panel: The reddest giants
are concentrated in a region in the south of the cluster field}
\end{figure}

\begin{table*}[t]
\caption[]{CN strengths (Norris \& Freeman 1982, 1983) and Str\"omgren colours of red giants in M22}
\begin{flushleft}
\begin{tabular}{lcccrrclcccrr}
\hline\noalign{\smallskip}
Nr.$_{\rm Arp}$ & $V$ & $(b-y)$ & $m_{1}$ & $\Delta m_{1}$ & S(3839) & & Nr.$_{\rm Arp}$ & $V$ & $(b-y)$ &
$m_{1}$ & $\Delta m_{1}$ & $S(3839)$\\
\noalign{\smallskip}
\hline\noalign{\smallskip}
I8   & 11.979 & 0.951 & 0.212 & $+$0.025 & 0.49 & & II67  & 11.253 & 1.239 & 0.506 & $+$0.141 & 0.64\\
I11  & 12.765 & 0.874 & 0.239 & $+$0.099 & 0.54 & & II77  & 12.677 & 0.935 & 0.068 & $-$0.109 & 0.14\\
I12  & 11.640 & 0.999 & 0.173 & $-$0.044 & 0.16 & & II80  & 11.488 & 1.149 & 0.420 & $+$0.111 & 0.85\\
I36  & 11.902 & 0.961 & 0.120 & $-$0.073 & 0.04 & & II92  & 12.555 & 0.932 & 0.242 & $+$0.067 & 0.76\\
I37  & 11.942 & 0.969 & 0.148 & $-$0.051 & 0.16 & & II98  & 12.274 & 0.927 & 0.164 & $-$0.008 & 0.23\\
I45  & 12.504 & 0.886 & 0.093 & $-$0.055 & 0.05 & & II104 & 12.322 & 0.959 & 0.124 & $-$0.068 & $-$0.01\\
I51  & 12.735 & 0.890 & 0.081 & $-$0.068 & $-$0.03 & & III3 & 11.130 & 1.175 & 0.563 & $+$0.238 & 0.50\\
I53  & 12.645 & 0.910 & 0.178 & $+$0.016 & 0.54 & & III6  & 12.786 & 0.896 & 0.086 & $-$0.067 & 0.14\\
I54  & 12.770 & 0.878 & 0.198 & $+$0.055 & 0.45 & & III12 & 11.503 & 1.106 & 0.452 & $+$0.170 & 0.58\\
I57  & 11.856 & 0.793 & 0.344 & $+$0.255 & 0.48 & & III14 & 11.096 & 1.201 & 0.445 & $+$0.104 & 0.00\\
I68  & 12.490 & 0.882 & 0.049 & $-$0.096 & $-$0.05 & & III15 & 11.325 & 1.106 & 0.403 & $+$0.120 & 0.33\\
I80  & 12.506 & 0.900 & 0.251 & $+$0.095 & 0.85 & & III25 & 12.640 & 0.850 & 0.026 & $-$0.099 & $-$0.14\\
I85  & 12.438 & 0.867 & 0.065 & $-$0.070 & 0.05 & & III26 & 11.111 & 1.151 & 0.362 & $+$0.051 & 0.06\\
I86  & 12.284 & 0.969 & 0.103 & $-$0.095 & 0.05 & & III35 & 12.353 & 0.930 & 0.098 & $-$0.076 & 0.06\\
I92  & 11.522 & 1.050 & 0.232 & $-$0.016 & 0.08 & & III39 & 12.533 & 0.898 & 0.095 & $-$0.059 & 0.24\\
I98  & 12.286 & 0.868 & 0.089 & $-$0.047 & 0.21 & & III45 & 12.818 & 0.869 & 0.136 & $-$0.001 & 0.40\\
I105 & 12.135 & 0.932 & 0.113 & $-$0.063 & 0.18 & & III47 & 12.329 & 0.904 & 0.128 & $-$0.030 & 0.19\\
I106 & 12.431 & 0.885 & 0.090 & $-$0.057 & 0.19 & & III52 & 11.556 & 1.081 & 0.549 & $+$0.282 & 0.55\\
I108 & 12.782 & 0.799 & 0.018 & $-$0.075 & $-$0.08 & & III61 & 11.791 & 1.012 & 0.259 & $+$0.034 & 0.50\\
I109 & 12.328 & 0.904 & 0.114 & $-$0.045 & $-$0.04 & & III86 & 12.439 & 0.885 & 0.156 & $+$0.009 & 0.32\\
I110 & 12.757 & 0.875 & 0.088 & $-$0.053 & 0.12 & & IV17  & 11.356 & 1.237 & 0.289 & $-$0.075 & 0.23\\
I113 & 12.375 & 0.938 & 0.096 & $-$0.083 & 0.29 & & IV20  & 12.041 & 1.021 & 0.336 & $+$0.106 & 0.72\\
I116 & 12.783 & 0.834 & 0.023 & $-$0.092 & 0.25 & & IV24  & 12.584 & 0.912 & 0.281 & $+$0.118 & 0.98\\
II1  & 12.032 & 1.003 & 0.321 & $+$0.101 & 0.50 & & IV25  & 12.842 & 0.819 & 0.044 & $-$0.062 & $-$0.03\\
II9  & 12.780 & 0.888 & 0.178 & $+$0.029 & 0.58 & & IV31  & 12.554 & 0.885 & 0.201 & $+$0.054 & 0.68\\
II23 & 12.596 & 0.876 & 0.092 & $-$0.049 & 0.09 & & IV59  & 11.875 & 1.002 & 0.144 & $-$0.074 & 0.23\\
II26 & 11.326 & 1.157 & 0.327 & $+$0.013 & 0.29 & & IV61  & 12.599 & 0.957 & 0.084 & $-$0.107 & 0.14\\
II30 & 12.147 & 0.942 & 0.144 & $-$0.037 & 0.12 & & IV67  & 12.389 & 0.905 & 0.060 & $-$0.099 & 0.03\\
II31 & 11.878 & 1.015 & 0.210 & $-$0.017 & 0.12 & & IV73  & 12.901 & 0.864 & 0.059 & $-$0.074 & $-$0.06\\
II42 & 12.767 & 0.865 & 0.072 & $-$0.062 & $-$0.01 & & IV76  & 12.257 & 0.916 & 0.289 & $+$0.123 & 0.79\\
\noalign{\smallskip}
\hline
\end{tabular}
\end{flushleft}
\end{table*}

Dividing the red giants in a red and a blue sample, as indicated in the upper
panel of Fig.~6 by filled dots and open triangles, respectively,
provides a simple means to assess the role of differential reddening. The
lower panel shows the distribution projected on the sphere. The redder stars 
clearly tend to be on the southern side, especially along a narrow region from
the center towards the south of the cluster. 
In contrast, many of the bluer giants are located on the north-eastern side.
Subdividing the cluster into quarters we find an excess of red stars in the
south-eastern quarter with a significance of 2$\sigma$. In contrast, there
is a deficiency of red stars (at a 4$\sigma$ level) in the north-eastern quarter
(see Fig.\,6, lower panel).
With regard to the cumulative distribution of red and blue stars in
the azimuthal angle 
we obtain a probability of
$\sim 95$ percent that the red and the blue giants are taken from a
different distribution function (Kolmogorov-Smirnov test).
This inhomogeneous 
distribution of the red and blue cluster giants indicates variations
in the reddening, obviously caused by patchily structured dust in the
foreground of M22. From the width of the giant branch (Fig.~6, upper panel) we
estimate a total range of $0.07$ mag for the reddening variations, after
taking a photometric error in $(b-y)$ of $0.019$ mag into account.

Fig.~7 provides an assessment of the appearance of ``red'' and ``blue''
red giants in the $m_1,(b-y)$ diagram, but no correlation between
the scatter in this diagram and the colour spread
in the CMD is visible. We would expect that ``red'' RGB stars show
systematically lower metallicities, if reddening would be dominantly
responsible for a stellar locus in the $m_1$,$(b-y)$ diagram.
Variable CN-abundances remain as a plausible explanation for the scatter.

\begin{figure}
\resizebox{\hsize}{!}{\includegraphics{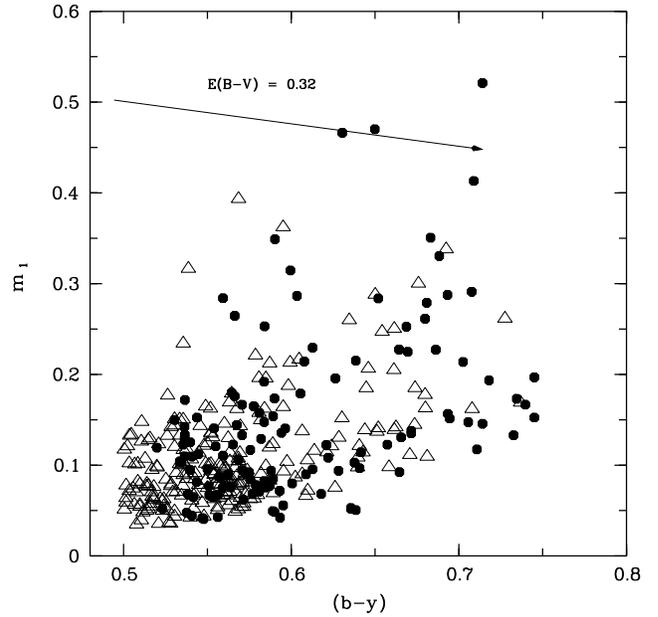}}
\caption{No correlation between the colour of giant branch stars 
(symbols are from Fig. 6)  and
the dispersion of $m_1$ values can be seen in the $m_1,(b-y)$ diagram
of M22. Therefore reddening variations can not be responsible for
the large scatter seen in this diagram}
\end{figure}

\subsection{CN abundances}
The strong CN-band at 4216 \AA\ can reduce the flux in the $v$ filter
significantly and thus may lead to a higher $m_1$ value, while
the overall metallicity remains the same (Bell \& Gustafsson 1978).
The distribution in the $m_1,(b-y)$ diagram therefore may be an
indication that the M22 giants exhibit strongly variing CN-bands.
A large range of CN abundances in M22 has indeed been 
spectroscopically observed by NF.

60 giant stars from the sample of NF with known CN abundances
have been identified and compared  with our data. Table 5 shows
these stars with the identification numbers from Arp \& Melbourne (1959),
magnitudes and colours from our photometry, and the CN band abundances
from NF, given by the index $S(3839)$. Values for the $S(3839)$ index vary
between $-0.14$ (CN weak) and $+0.98$ (CN strong) and the CN variations
therefore are significantly higher than observed in M55. To investigate the
influence of the CN strengths on the $m_1$ index in the case of M22,
all CN strong stars with $S(3839)$ values larger than $+0.40$
have been plotted in the $m_1,(b-y)$ diagram (Fig.~8, upper panel) as filled circles,
while the CN weak stars ($S(3839) \le 0.40$) are shown as open circles. 
The correlation between the positions of the 
giants in the $m_1,(b-y)$ diagram and their
CN abundances is clearly visible. All stars with strong CN abundances
are located in the upper part of the diagram and it is reasonable
to conclude that also all the other stars in this area, that have no
spectroscopic measurements, are CN rich as well. 

The influence of the cyanogen strengths on the
$m_1$ index can be seen in more detail by using the metallicity calibration for the 
$m_1,(b-y)$ diagram. The mean Str\"omgren metallicity (using the
calibration of Hilker) of all 60 giant
stars with known CN abundances is $-1.73$ dex,
taking a reddening value of $E(B-V)=0.32$ into account (the CN strengths
definitely influence this value; it is therefore too high). 
In Fig.\,8 (lower panel), $\Delta m_1$ (as defined in Sect.~3.3) 
to the calibration line of $-1.73$ dex is plotted versus $S(3839)$. In contrast to M55,
the influence of the cyanogen strenghts on the $m_1$ index in M22 is 
significant. $\Delta m_1$ becomes larger by increasing CN abundances and
the CN variations therefore shift the giant stars to higher $m_1$ values.
This effect dominates the shape of the $m_1,(b-y)$ diagram, it is even stronger
than the reddening variations of about $0.07$ mag.
Our results are similar to investigations of Anthony-Twarog et al. (1995),
who investigated CN and Ca abundances in relation with the Str\"omgren
$m_1$ index for M22. 
We compared 15 stars from their observations with our data stars and find 
a good agreement in the measured Str\"omgren colours. Mean deviations are 
$\Delta V =-0.029$, $\Delta (b-y)=-0.034$
and $\Delta m_1 =-0.010$ (see also Sect.\,3.3).
 
The fact that only strong cyanogen abundances
influence the $m_1$ index offers us the possibility to determine 
the mean Str\"omgren metallicity and the mean reddening for M22 
using only the 40 spectroscopically measured
CN weak giants. With the method described in Sect.\,3.2  we derive a cluster metallicity of
$-1.62 \pm 0.08$ dex and a reddening of $E(B-V)=0.38 \pm 0.04$.
As for M55, the derived metallicity of M22 is (with regard to the 
expected errors) in agreement with recent spectroscopic results, such as
from Carretta \& Gratton (1997, $-1.48 \pm 0.06$ dex) and Lehnert et al. 
(1991, $-1.55 \pm 0.11$ dex).   

\begin{figure}
\resizebox{\hsize}{!}{\includegraphics{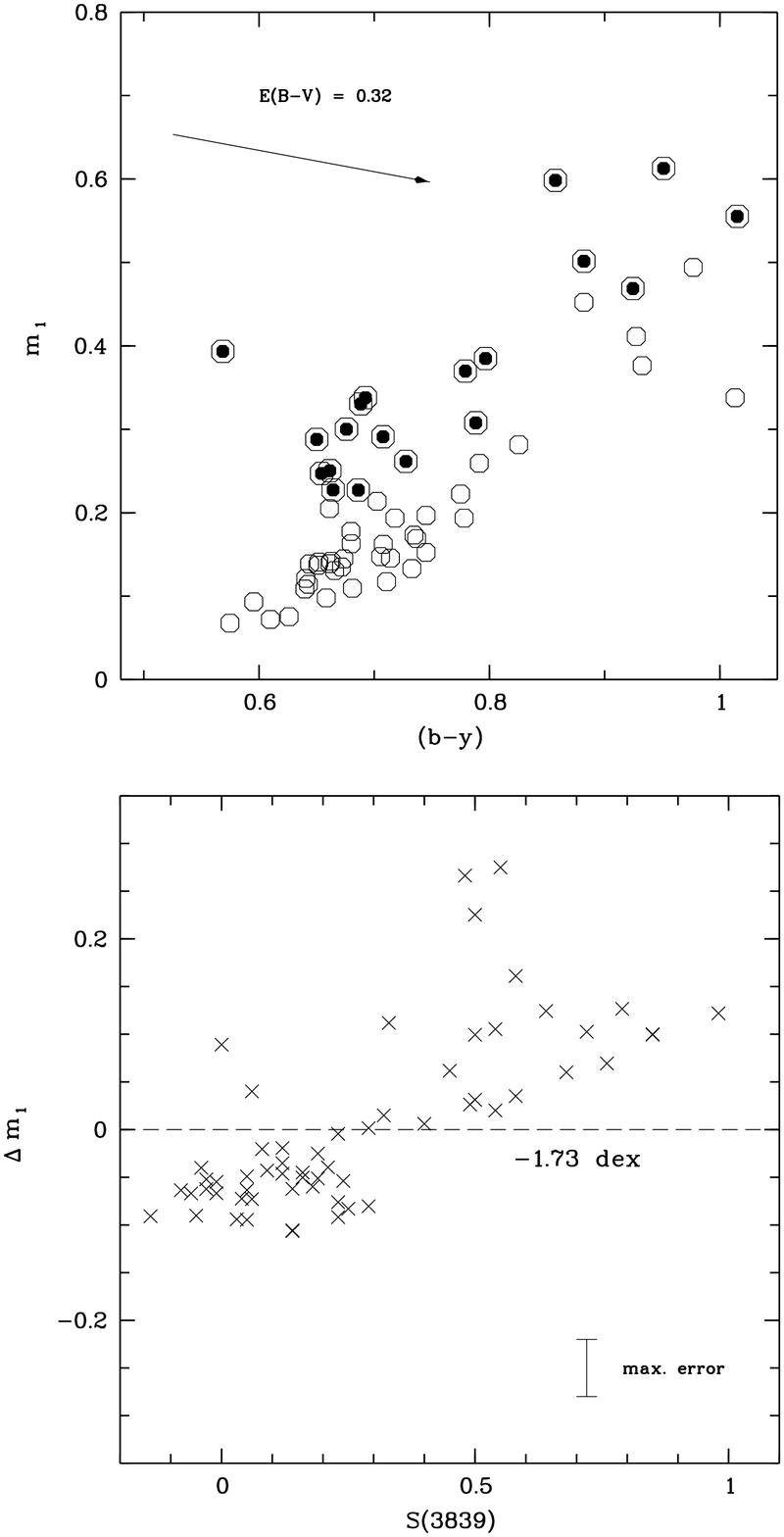}}
\caption{Upper panel: All CN rich stars in M22 show higher $m_1$ values
than comparable stars with lower CN abundances. Lower panel: The CN
abundances are correlated to the vertical distance of
the slope which represents the mean Str\"omgren metallicity
for the selected giants in M22 with an adopted reddening of $E(B-V)=0.32$}
\end{figure}

\section{Discussion and conclusions}
A comparative study of the two galactic globular clusters M55 and M22 has 
shown that the Str\"omgren system can efficiently detect differences in the CN-strength
distribution of globular clusters. 
Moreover, with the method presented in Sect.\,3.2 Str\"omgren photometry 
can be used for the 
simultaneous determination of mean metallicities {\it and} reddening values
in globular clusters.

The present study of M55 and M22 has revealed that the clusters are 
very different regarding the distribution of Str\"omgren
colours.
M55 turns out to be very uniform with respect to Str\"omgren-metallicities.  
The sharp red giant branch
in the CMD as well as in the $m_1,(b-y)$ diagram indicates the chemical
homogeneity of this globular cluster. 

For M22, we confirm the results by Anthony-Twarog et al. (1995).  
A large scatter in the Str\"omgren $m_1$,$(b-y)$ diagram indicates a large spread in
CN strengths among red giants.
Therefore, Str\"omgren colours of M22 stars cannot be interpreted in terms
of an overall metallicity, in particular nothing can be said concerning
an intrinsic metallicity variation among M22 stars.  
Moreover, the colour spread of the RGB and
the BHB in the colour-magnitude diagram of M22 is not caused by a metallicity
variation but probably by patchy
reddening variations in the order of $\Delta E(B-V) \approx 0.07$ over
the cluster area, because most of the reddest giant-branch stars 
are spatially concentrated in the southern part of the cluster.
This, and the fact that there is no correlation between the 
dispersions in the CMD and the $m_1,(b-y)$ diagram,
led us to conclude that a spread in [Fe/H], if present, is of minor importance
with respect to the CN scatter.
Considering similar results for M22 of Anthony-Twarog et al. (1995)
and the spectroscopic measurements by Lehnert et al. (1991),
$\omega$ Cen remains as the only globular cluster in the Milky
Way, where significant evidence exists for a possibly primordial dispersion in [Fe/H].

So far, the existence of a wide range of cyanogen strengths 
has been shown only in a few globular clusters (see Kraft 1994 for a detailed
overview) and their origin can be explained by primordial as well as by 
evolutionary origin. See to this also the extensive discussion in Anthony-Twarog
et al. (1995).  
Perhaps, CN-rich clusters may be brought into relation with anomalies found
in the comparison of integrated spectra of M31 clusters with those of
galactic globular clusters. Several authors found (Burstein et al. 1988; Brodie \& Huchra 1990) that, at the same
metallicity, M31 clusters showed stronger CN-features than galactic clusters.
Unfortunately, no work exists measuring integral CN strengths for both M22 and
M55.

It is of obvious interest to perform a larger census of the CN-strength distribution
among globular-cluster stars. The Str\"omgren system proved to be an efficient
tool for this purpose. At present, it is not possible to investigate the relation
of ``CN-richness'' with any other properties of globular clusters. 
M55 and M22 are perhaps representatives of two classes of
globular clusters regarding their CN-strength distribution. If the CN-strengths in
M22 are decoupled from stellar evolution, what else can be the cause? A striking difference
between M55 and M22 is their stellar density. M55 is one of the most loosely
structured globular clusters while, in contrast,  M22 has a quite high density. It is imaginable that
encounters or merging of stars can change the CN-surface abundance of red giants.
It is therefore of high interest to study a larger sample of globular clusters.   

\acknowledgements
We would like to thank K.S. de Boer and T. Puzia for helpful 
comments on this work.

\appendix

\section{Blue stragglers in M55}

\begin{normalsize}
As mentioned in Sect. 3.1, a population of blue straggler stars (BSS) is
visible in the CMD of M55 in the area above the turn-off point.
The existence of BSS in M55 and their central concentration has already
been shown
in detailed analyses by Mandushev et al. (1997) and Zaggia et al. (1998).

\begin{figure}
\resizebox{\hsize}{!}{\includegraphics{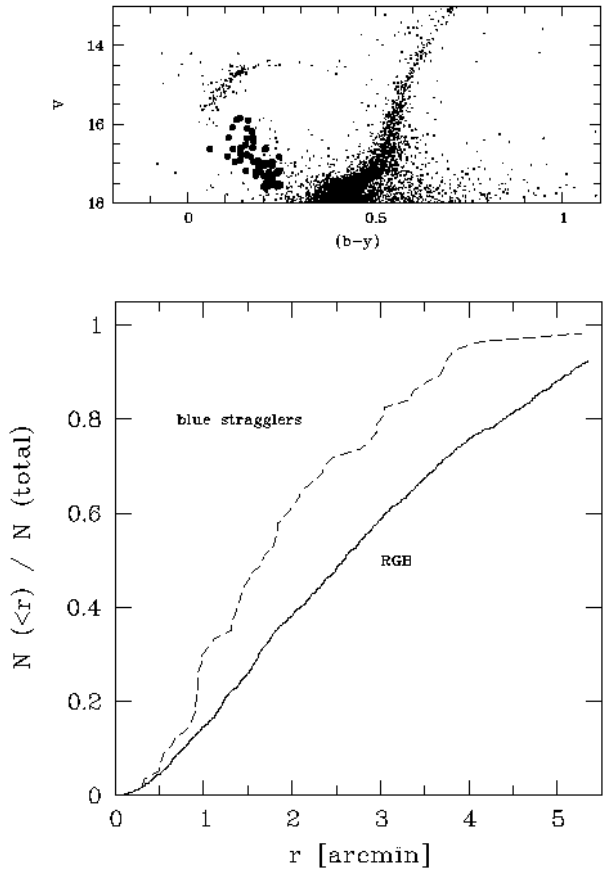}}
\caption{Upper panel:  The 56 BSS in the CMD of M55.
Lower panel: The cumulative
distribution show that the BSS are more centrally concentrated
than RGB stars}
\end{figure}

At present there is much evidence that BSS are close or merged
binary stars (e.g. see Mandushev 1997 and references therein).
This theory is supported by the finding that the BSS 
are often more centrally concentrated than other stellar types
within globular clusters, which might be indicative for a dynamical
mass segregation. In the CMD the BSS sequence ranges
up to 2.5 magnitudes above the TOP (see Fig.\,A1).
From our data sample of M55 we selected 56 BSS 
candidates from the inner $5\farcm3$ for
further investigations, shown in Fig.\,A1 as filled circles.

We compared the spatial distribution of the 56 BSS with that of 1377
sub-giant branch (SGB) stars chosen from the same magnitude interval
to ensure that incompleteness does not influence the comparison. In Fig.\,A1,
the cumulative distributions of the SGB stars and the BSS are plotted.
The BSS are clearly more concentrated than the SGB stars within the
selected radius. Running a Kolmogorov-Smirnov test we obtain a 92 percent
probability that the BSS and the SGB stars are taken from different
distributions, a value that is similar to the one found by
Mandushev et al. (1997) for M55. These results favour mass
segregation and further support that BSS have on the average larger masses
than subgiants and thus are merged stars or binaries.

\end{normalsize}

\end{document}